\documentclass[aps,twocolumn,prl,groupedaddress,amsmath,runinaddress,showpacs]{revtex4}
\usepackage{graphicx}
\usepackage{subfigure}
\renewcommand{\vec}[1]{\mathbf{#1}}

\begin{document}
\title{How spherical plasma crystals form}
\author{H. K\"{a}hlert}
\author{M. Bonitz}
\affiliation{Institut f\"{u}r Theoretische Physik und Astrophysik, Christian-Albrechts Universit\"{a}t zu Kiel, 24098 Kiel, Germany}


\date{\today}

\begin{abstract}
The correlation buildup and the formation dynamics of the shell structure in a spherically confined one-component plasma are studied.
Using Langevin dynamics simulations the relaxation processes and characteristic time scales and their dependence on the pair interaction and 
dissipation in the plasma are investigated.
While in systems with Coulomb interaction (e.g. trapped ions) in a harmonic confinement shell formation starts at the plasma edge and proceeds inward, this trend is significantly weakened for dusty plasmas with Yukawa interaction. With a suitable change of the confinement conditions 
the crystallization scenario can be externally controlled.
\end{abstract}


\maketitle

Crystallization of charged particles, predicted by Wigner seven decades ago, continues to stimulate research in many fields due to its relevance for astrophysics (e.g. white dwarf stars), basic many-body physics and potential applications in quantum computing, e.g. \cite{drewsen09}. Experimental realizations include electrons on a helium surface \cite{rousseau09}, ion in traps \cite{itano1998,drewsen09}, electrons in quantum dots \cite{filinov_prl01}, dusty plasmas \cite{dustchina,thomas_prl94} and ultracold neutral plasmas ~\cite{killian2007,rolston2008}. For crystallization to be possible in thermodynamic equilibrium in a macroscopic three-dimensional one-component plasma (OCP), the coupling parameter, $\Gamma=q^2/(a_{\text{ws}}k_\text{B} T)$ has to exceed a value of about $174$, where $a_{\text{ws}}$ denotes the Wigner-Seitz radius, $q$ the charge, $T$ the temperature and $k_\text{B}$ Boltzmann's constant, whereas additional conditions have to be fulfilled in two-component plasmas (TCP) \cite{bonitz-prl05}. Recently, crystallization of spherically trapped dust particles has been achieved \cite{arp2004} which revealed close similarities to ion crystals in traps \cite{itano1998,drewsen09} with the main difference being the screening of the Coulomb interaction in the former case \cite{bonitz2006,henning08}.

While the stuctural properties of the crystals are well understood, e.g. \cite{schiffer03,bonitz-prl05}, much less is known on the dynamics of their formation. Murillo showed \cite{murillo2001} that a neutral TCP produced by rapid ionization of an atomic gas will not crystallize because the correlation buildup is accompanied by heating \cite{bonitz97} which limits $\Gamma$. Then Pohl \emph{et al.} demonstrated that crystallization can be achieved if the expanding plasma is laser-cooled~\cite{pohlcryst}, which still has not been realized experimentally \cite{killian2007,rolston2008}. An interesting prediction of \cite{pohlcryst} was that spherical crystal shells {\em start to form in the cluster core}. It is an open question whether this is a general crystallization scenario in trapped plasmas since, so far, no investigations on the crystal formation dynamics in spherically trapped ions and dusty plasmas have been performed.

The goal of this Letter is, therefore, to perform such an analysis for spherically confined dusty plasmas. We study in detail the time-dependence of crystallization by simulating an experimental cooling process from a weakly correlated finite dust cloud towards strong coupling. We show that the formation of spatial correlations proceeds in a sequence of stages and present results for the characteristic time scales. Further, the dependence of the dynamics on  screening and dissipation is explored. We predict that the onset of shell formation is typically at the cluster edge but the order of appearence of the inner shells can be controlled by suitable variation of the confinement. Finally, when the core region is made unaccessible to the plasma crystallization 
can be initiated in the center.

\emph{Model and simulation idea.} We consider $N$ identical particles with mass $m$ and charge $q$ interacting through a Yukawa pair potential~\cite{bonitz2006} $\phi(r)=q^2e^{-\kappa r}/r$ in an external confinement $V(r)$. The effective range of $\phi(r)$ is determined by the screening parameter $\kappa$. The dynamics of our system is described by the Hamiltonian
\begin{equation}\label{eqn:Hamiltonian}
  H=\sum_{i=1}^{N}\frac{\vec{p}_ i ^2}{2m}+\underbrace{\sum_{i=1}^{N} V(|\vec{r}_i|) + \frac{1}{2}\sum_{i\ne j}^N \phi(|\vec{r}_{ij}|)}_{U(\vec r_1,\ldots,\vec r_N)}.
\end{equation}
With a harmonic confinement $V(r)=m\omega_0^2r^2/2$ this model accurately describes the properties of spherical dust balls observed in recent experiments~\cite{bonitz2006} and is equally applicable to spherically trapped ions
 in the limit $\kappa=0$. Below we use dimensionless units with the characterstic length and energy scales $a=(q^2/m\omega_0^2)^{1/3}$ and $E_0=q^2/a$.

The ambient neutral gas in dusty plasma experiments is accounted for by an additional damping term and a fluctuating force in the (Langevin) equation of motion of the $i$-th particle
\begin{equation}\label{eqn:LangevinEQ}
m\ddot{\vec r}_{i}=-\nabla_i U(\vec r_1,\ldots,\vec r_N)-m\nu \dot{\vec r}_i+\vec{f}_i(t).
\end{equation}
The friction coefficient $\nu$ and the Gaussian noise $\vec f _i(t)$ are related by the fluctuation-dissipation theorem $\left< \vec{f}_i^{\alpha}(t) \vec{f}_j^{\beta}(t') \right>=2 m \nu k_\text{B} T^* \delta_{ij} \delta_{\alpha\beta} \delta (t-t')$, where $\alpha,\beta\in\{x,y,z\}$ and $i,j\in\{1,\ldots,N\}$. In Ref.~\cite{pohlcryst} this method was used to describe the cooling effect of the laser.

We consider the following scenario to study the buildup of correlations: We start from a weakly correlated steady state of $N$ trapped dust particles characterized by $\Gamma_\text{i}=q^2/(a\, k_\text{B} T_\text{i})=0.2$ which can be prepared e.g. by continuous laser-heating, which has been successfully applied in experiments on 2D layers~\cite{laserheating}. 
At the time $t=0$ the laser is turned off and the particles begin to slow down at a rate determined by the friction coefficient $\nu$, approaching a new equilibrium at the neutral gas temperature $T_\text{n}$ corresponding to strong coupling with $\Gamma_\text{n}=125$. 
This scenario allows us to study the correlation buildup in a well defined manner without introducing spatial inhomogeneities. The chosen $T_\text{n}$ is of the order of recently measured normal mode temperatures~\cite{ivanov3d}.
Due to the external trap the system is inhomogeneous but isotropic, i.e. the mean density will only depend on the distance $r$ from the trap center. To reduce  the effect of fluctuations we perform several hundred to $1200$ runs with random initial conditions over which we average. 

%
\begin{figure}[h]
 \centering
 \includegraphics[width=0.49\textwidth]{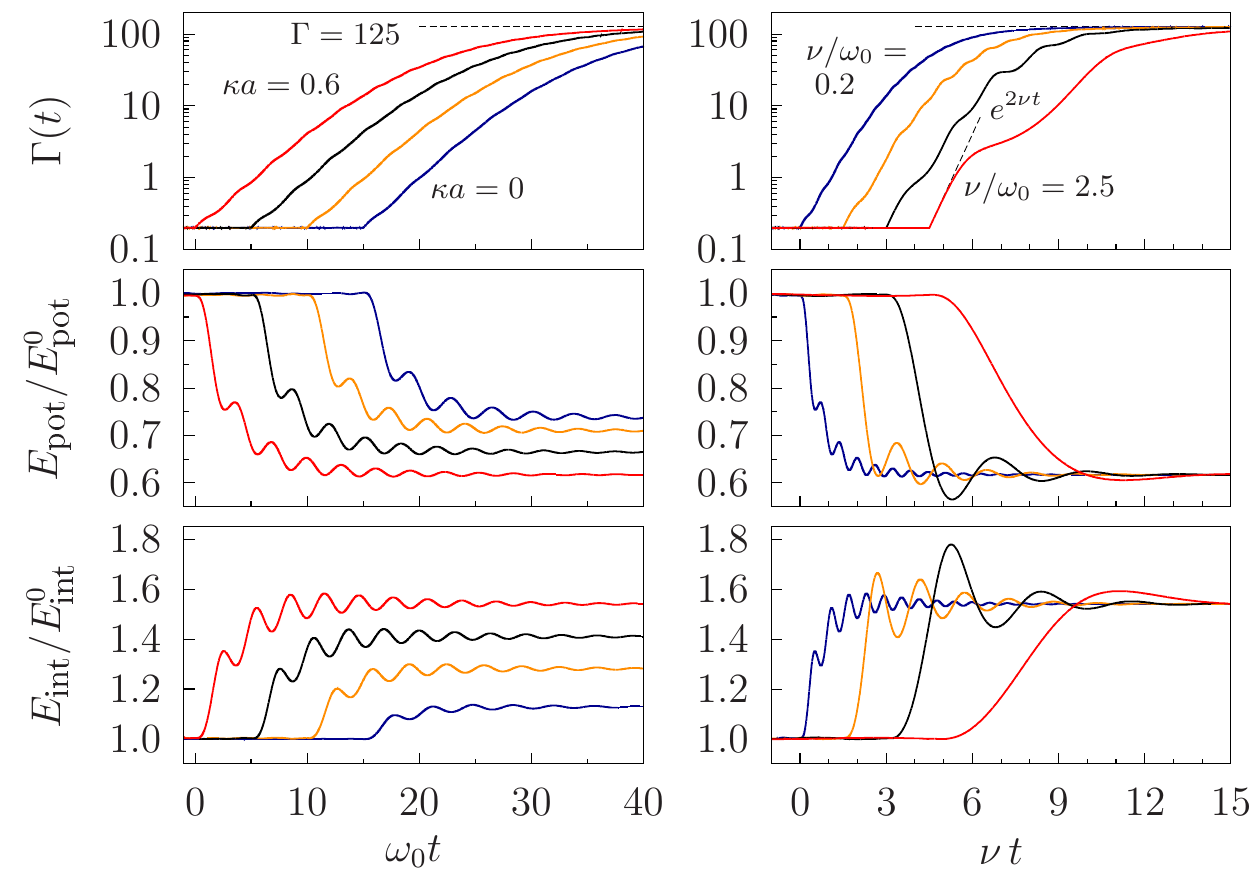}
 \caption{(Color online) Time dependence of the kinetic, potential and interaction energy for $N=400$. For the sake of clarity the graphs for different parameters are shifted by $\Delta(\omega_0 t)=5$ (left) and $\Delta(\nu t)=1.5$ (right). Left: Influence of screening for $\nu/\omega_0=0.2$ and $\kappa a=0.6,\,0.4,\,0.2,\,0$ (from left to right). Right: Influence of the damping rate for $\kappa a=0.6$ and $\nu/\omega_0=0.2,\,0.5,\,1,\,2.5$. Note the different scaling of the time axes. Potential and interaction energy are normalized to the equilibrium energies at $\Gamma_\text{i}=0.2$.}\label{fig:energy}
\end{figure}
\emph{Cooling towards strong coupling.} Consider first the evolution of the coupling parameter $\Gamma(t)$ which we compute from an instantaneous 
temperature $k_\text{B}T(t)=2 E_{\text{kin}}(t)/3N$ \cite{temperature}. Fig.~\ref{fig:energy} shows that $\Gamma(t)$ increases continuously reaching 
the value $125$ within $\omega_0 t \approx 35$ \cite{crystal}. The increase of $\Gamma(t)$ is accompanied by weak oscillations and is only marginally affected by $\kappa$, cf. left column. In contrast, the influence of friction is more apparent: an increase of $\nu$ leads to an increase of the modulation amplitude, cf. right column of Fig.~\ref{fig:energy}. For strong friction the initial growth follows $\Gamma(t)\propto\exp(2\nu t)$, which is the expected behavior for a free particle subject to friction (ballistic regime).

Fig.~\ref{fig:energy} also shows a non-trivial dynamics of the confinement ($E_{\rm pot}$) and interaction energy ($E_{\rm int}$) during the crystallization. While $E_{\rm pot}$ decreases, due to compression of the cluster, $E_{\rm int}$ inreases due to the formation of correlations. The relative gain (loss) of interaction (confinement) energy increases with screening.
%
%
$E_{\rm pot}$ and $E_{\rm int}$ exhibit much more pronounced oscillatory modulations than $\Gamma(t)$ which are determined by dissipation. While for $\nu/\omega_0=0.2$ only small oscillations occur, for $\nu/\omega_0=0.5$ an overshooting of $E_{\rm pot}$ is observed which reaches its maximum 
for $\nu/\omega_0=1$. Upon further increase of $\nu$ the oscillations vanish.
\begin{figure}
 \centering
 \includegraphics[width=0.49\textwidth]{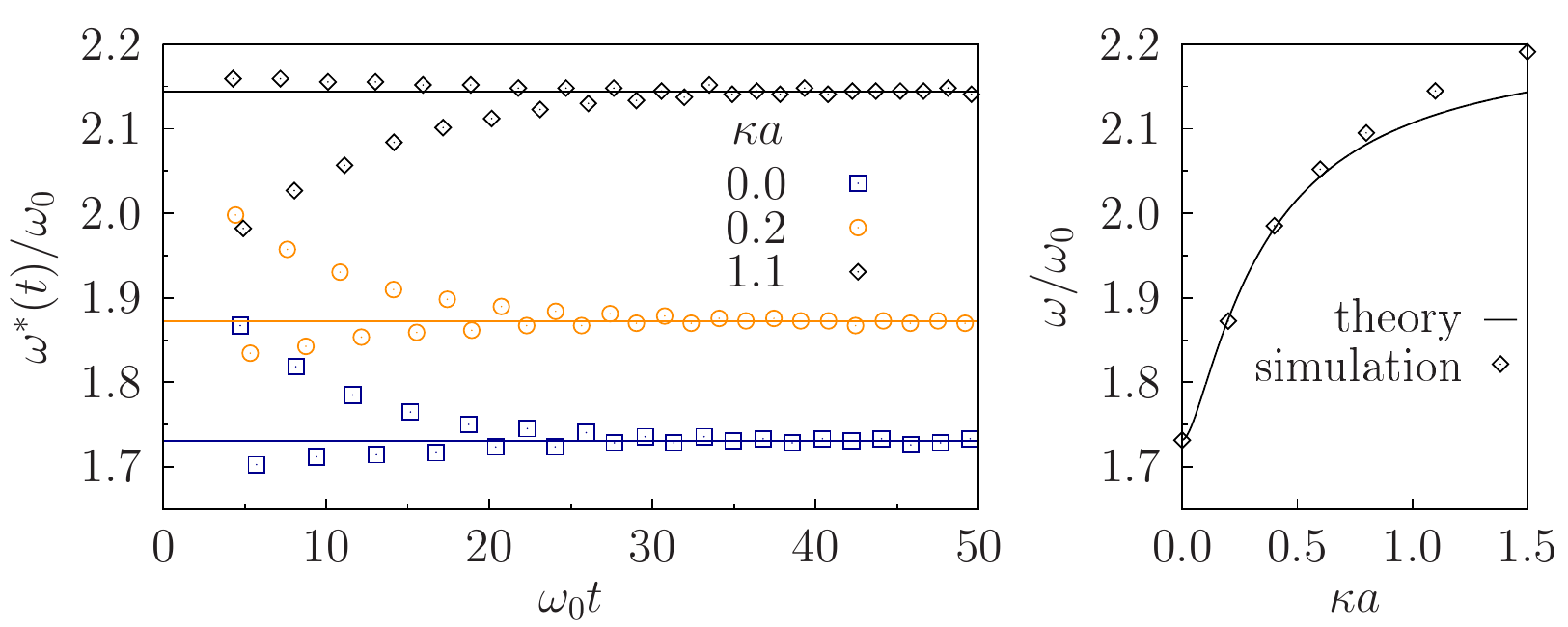}
 \caption{(Color online) Left: Instantaneous oscillation frequency of $E_{pot}$ (cf. Fig.~\ref{fig:energy}) for various $\kappa$ at $\nu/\omega_0=0.1$. Horizontal lines denote the mean frequency in the interval $35\le \omega_0 t\le 50$. Right: Mean frequency (corrected for friction, see text) compared to the analytical expression of Ref.~\cite{sheridan2006}.}\label{fig:frequency}
\end{figure}

The origin of the oscillations is easy to understand. When the heating is turned off the amplitude of the random force is reduced by $\Delta f_0=\sqrt{\nu_i T_i} - \sqrt{\nu T_n}$, giving rise to a rapid radial contraction of the cluster which excites a monopole oscillation. 
An increase of $\nu$ leads to a faster loss of kinetic energy and a stronger contraction, 
 explaining the larger oscillation amplitude. For $\nu \gtrsim 2 \omega_0$ the oscillation is overdamped and the amplitude decreases, whereas for $\nu\ll \omega_0$, $\Delta f_0$ is small and the system smoothly evolves from one equilibrium state to another. Therefore, there exists a maximum in the oscillation amplitude observed at $\nu\approx \omega_0$. Let us now analyze the oscillation frequency. To this end we compute an instantaneous frequency $\omega^*([t_i+t_{i+1}]/2)$ from two successive minima or maxima of the potential energy at $t_i$ and $t_{i+1}$, see left part of Fig.~\ref{fig:frequency}. After a few cycles the frequency saturates and the oscillations correspond to a damped normal mode of the new equilibrium state.
The intrinsic normal mode frequency $\omega$ (of the dissipationless system), right part of Fig.~\ref{fig:frequency}, is computed via $\omega^*=(\omega^2-\nu^2/4)^{1/2}$~\cite{henning09}.
For Coulomb interaction $\omega$ agrees with the breathing frequency $\omega_\text{br}=\sqrt{3}\,\omega_0$~\cite{henning08}, whereas in the case of finite screening it depends on $\kappa$. In the right part of Fig.~\ref{fig:frequency} we also display an analytical result for $\omega_\text{br}(\kappa R_0)$ derived for a homogeneous Yukawa sphere~\cite{sheridan2006} which is accurate at low screening. The normalized radius $\kappa R_0$ is computed from a mean-field theory~\cite{Christian2006,better_om_br}, see Eq.~(\ref{eqn:mf-density}) below. 

\begin{figure}
 \centering
 \includegraphics*[width=0.48\textwidth]{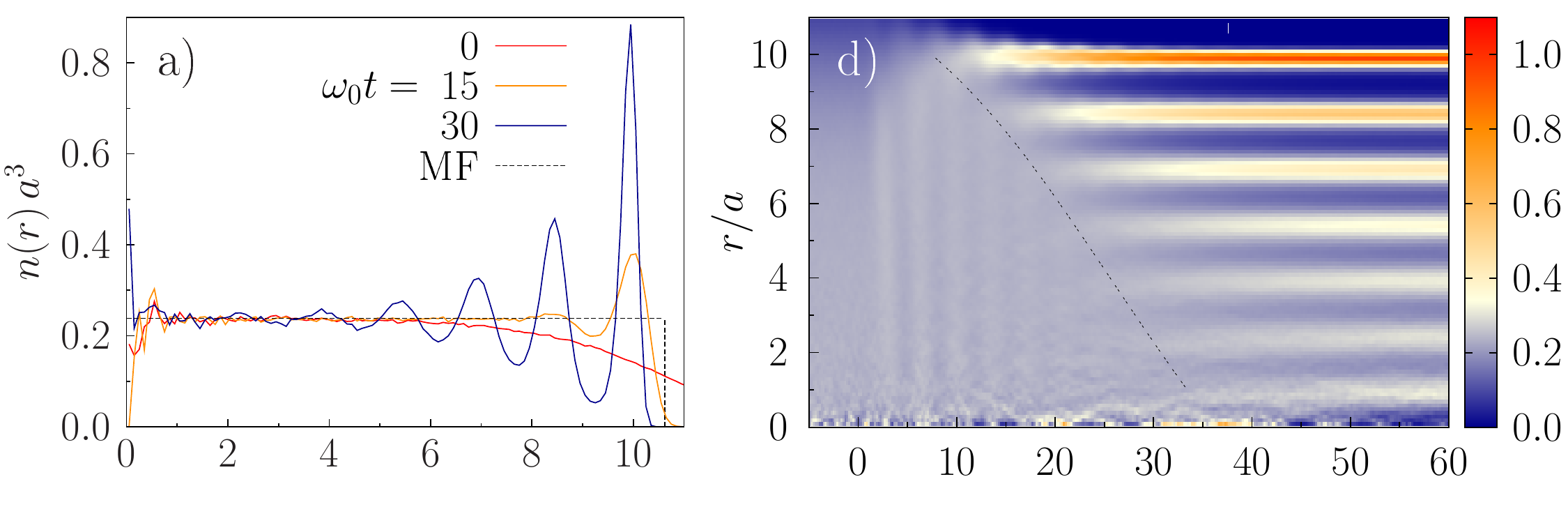}\\
 \includegraphics*[width=0.48\textwidth]{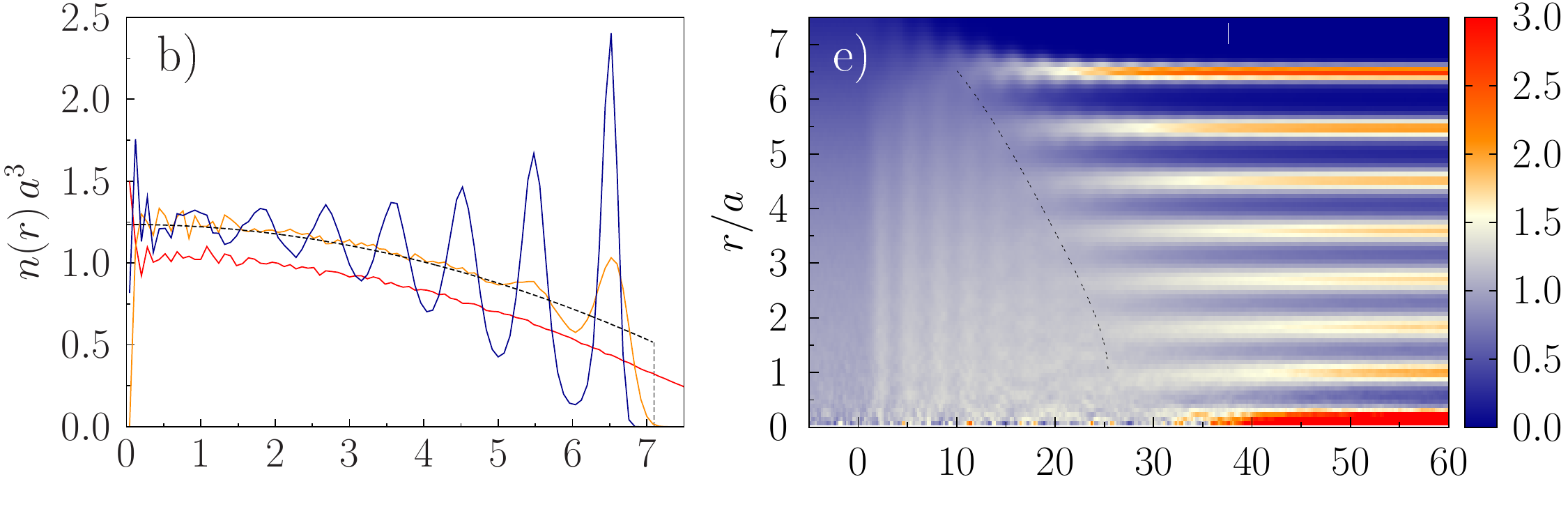}\\
 \includegraphics*[width=0.48\textwidth]{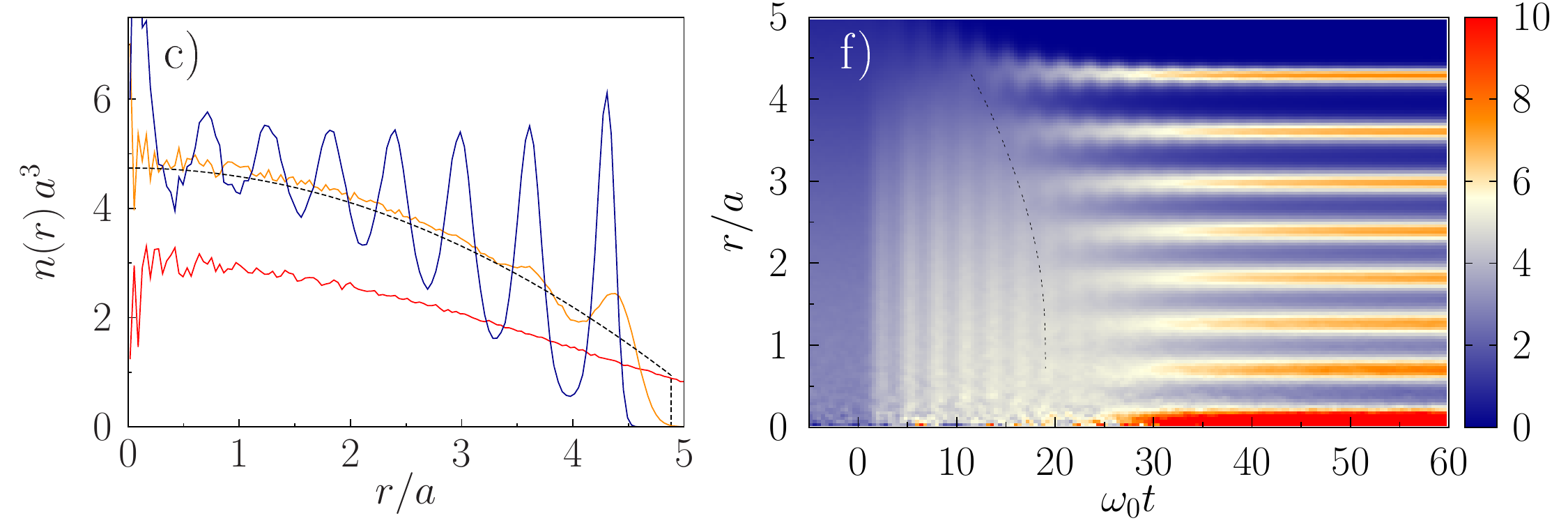}
 \caption{(Color online) Right: Evolution of density profile for $N=1200$ with $\nu/\omega_0=0.2$, black dashed line connects the approximate times and positions at which the shells emerge. Left: Snapshots of $n(r)$ at $\omega_0 t=0,\,15,\,30$ (oscillations grow with time) together with solution of Eq.~(\ref{eqn:mf-density}), ``MF'', black dots. From top to bottom row: $\kappa a=0;\, 0.6;\, 2$}\label{fig:density}
\end{figure}

\emph{Time-dependent density profile.}
The evolution of the radial density profile is shown in Fig.~\ref{fig:density}. In the initial weakly coupled state ($\Gamma_\text{i}=0.2$) the density is monotonically decaying and is well 
described by the Boltzmann factor, $n_{\rm eff}(r)\propto \exp[-V_{\rm eff}(r)/k_\text{B} T_\text{i}]$, where $V_{\rm eff}$ denotes the sum of confinement and mean field potential \cite{Christian2006}. As the kinetic energy drops the system approaches the strong coupling regime with its characteristic shell structure. In the case of Coulomb interaction (a, d) the first shell appears around the time $\omega_0 t\approx 10$, at the cluster boundary. Only upon further cooling [increase of $\Gamma(t)$] shells form one after another in the direction of the trap center after almost constant time intervals, cf. dashed line in Fig.~\ref{fig:density}.d.
The situation is different in the case of Yukawa interaction. While shell formation again starts at the edge, inner shells form more rapidly, cf. dashed line in Fig.~\ref{fig:density}.e, and at $\kappa a=2$, (Fig.~\ref{fig:density}.f), the inner shells emerge almost simultaneously.

This sequence of shell formation is in striking contrast to the one observed in expanding laser-cooled plasmas \cite{pohlcryst} where shells emerge in the center, which was attributed to an increased density in the core. In fact, the mean density profile $\bar{n}(r)$ allows one to define a local coupling  parameter $\bar{\Gamma}(r)\sim \bar{n}(r)^{1/3}$, and it is tempting to exepect shell formation to start at a radius where $\bar{\Gamma}(r)$ has its maximum.
Our results allow us to verify this hypothesis.  During the initial phase, $0\le \omega_0 t \lesssim 10$, $\bar{n}$ evolves from 
the Boltzmann factor, $n_{\rm eff}(r)$, to a (still) monotonous profile, cf. Fig.~\ref{fig:density}.a-c, which is very well described by the zero-temperature mean field result $\bar{n}_0$ given by~\cite{Christian2006} 
\begin{equation}\label{eqn:mf-density}
\bar{n}_0(r,N)=\left[\Delta V(r)-\kappa^2 V(r)+\kappa^2\mu(N)\right]/4\pi q^2,
\end{equation}
where $\mu(N)$ is the chemical potential, except for a slightly smoother decay at the edge. Note that Eq.~(\ref{eqn:mf-density}) exhibits a finite density step $\Delta n(R_0)$ at some maximum radius $R_0$ [$\bar{n}_0(r,N)\equiv 0$ for $r\ge R_0$] which emerges in our simulations rapidly, within $\omega_0 t \lesssim 10$.

In the Coulomb case [Fig.~\ref{fig:density}.a], $\bar{n}(r)$ is almost constant for $\omega_0 t \lesssim 10$, with a slight decay towards the edge, in agreement with Eq.~(\ref{eqn:mf-density}) which predicts a constant density for $r\le R_0=10.6$. Nevertheless, shells appear at very different moments starting at the edge, where the density is smallest. For $\kappa a=2$, the mean density decreases even stronger, $\bar{\Gamma}(R)/\bar{\Gamma}(0)\approx (1/2)^{1/3}$, cf. Fig.~\ref{fig:density}.c, but even here shells form at the edge first, in contrast to the above expectation. 
In fact, just prior to formation of the first shell around $\omega_0 t \sim 10$ the coupling parameter $\Gamma(t)$ approaches $10$, cf. Fig.~\ref{fig:energy}, where the mean-field description fails and correlations become important. With increasing $\Gamma(t)$ the discrete nature of the particles begins to manifest itself leading to formation of a (correlation) ``hole'' around each particle which cannot be occupied by others. This separation of particles in radial (and tangential) direction and an overall expansion of the cluster cause an increase of $E_{\rm pot}$ proportional to the number of particles in the edge layer, $N_{\rm edge}\sim \Delta n(R)$. The system reduces this energy by spontaneously restructuring such that particles from the edge are accumulated at a smaller distance -- the outer shell forms. The formation of the inner shells is triggered by the continuing increase of $\Gamma$ in the center. The substantial acceleration of shell formation with increasing $\kappa$ is explained by the inward force exerted by the particles on the outer shell~\cite{Christian2006} which is negligible in the Coulomb case (Faraday cage effect).

\begin{figure}
 \centering
 \includegraphics*[width=0.48\textwidth]{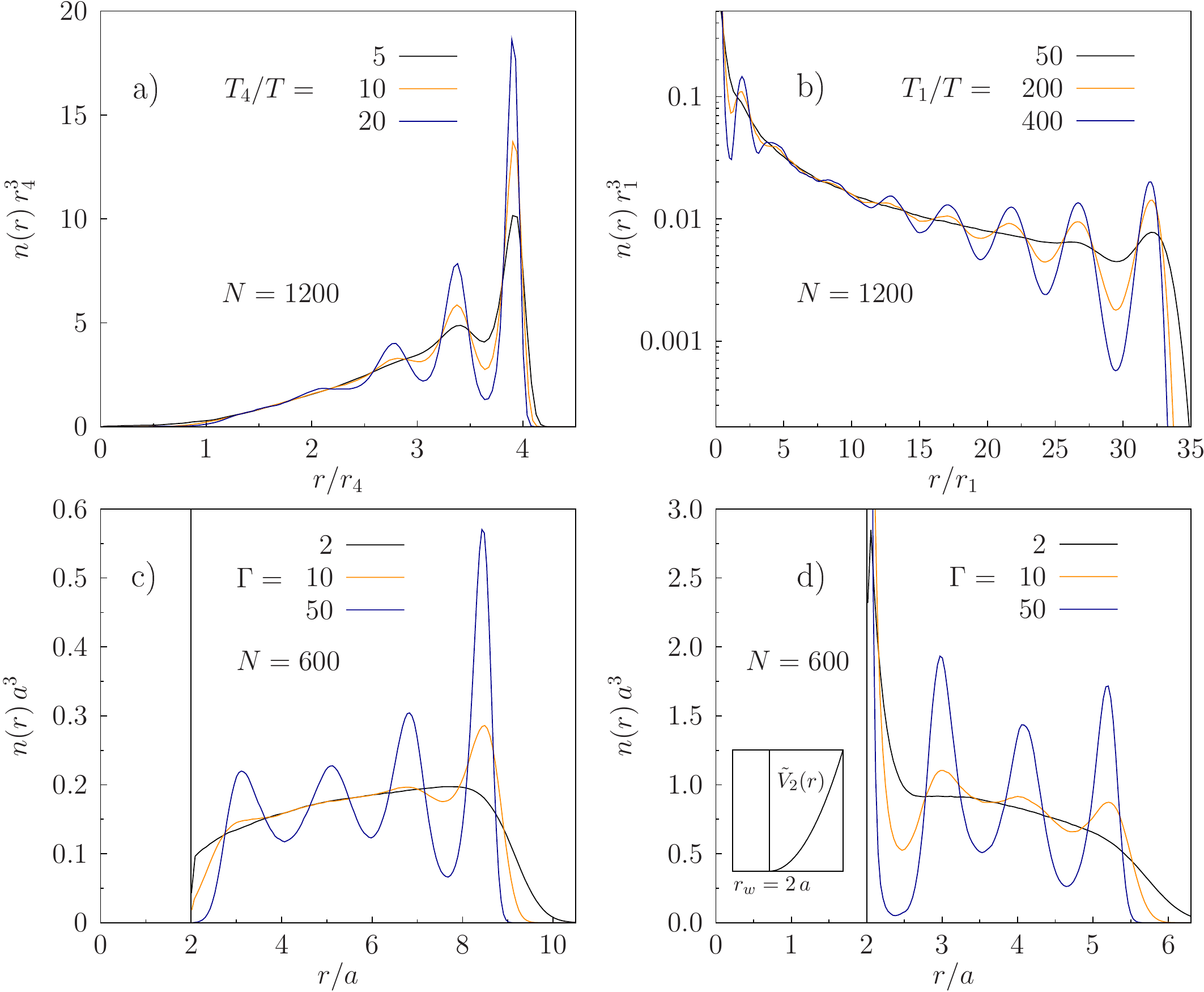}\\
 \caption{(Color online) Equilibrium density profiles for different trap potentials and various temperatures: a) quartic confinement $V_4$, b) 
linear confinement $V_1$. c) and d): harmonic potenital with blocked core, ${\tilde V}_2$, $r_w/a=2$. 
 Units are, $r_1= q/\sqrt{c_1}$, $r_4=\sqrt[5]{q^2/c_4}$ and $T_{\text{1,4}}=q^2/(r_{\text{1,4}}\, k_\text{B})$.}\label{fig:densityConf}.
\end{figure}

\emph{Effect of the confinement potential.} The different crystallization behavior observed in the expanding neutral plasma can be traced to a different confinement potential.
There, the ions are confined by the mean-field potential of the electrons~\cite{pohlcryst} which is Coulomb-like (except for the center), i.e. $V(r)\sim r^{-1}$. Consequently the mean density profile, prior to shell formation, exhibits a drastic 
increase towards the center, ${\bar n}(r) \sim r^{-3}$, cf. Eq.~(\ref{eqn:mf-density}), and a vanishing density step $\Delta n(R) \to 0$. The latter arises from vanishing of $V(r)$ for $r\to\infty$, i.e. ions with a finite kintic energy cannot be confined; instead of accumulating particles in a shell, the system expands. The strong increase of ${\bar n}(r)$ towards the center then explains the observed dynamics.
We now further verify the governing role of the confinement potential for the dynamics of shell formation. To this end we analyze the equilibrium density profile \cite{schiffer} for different temperatures by performing thermodynamic Monte Carlo simulations for a finite number of particles trapped by different potentials.
According to Eq.~(\ref{eqn:mf-density}) a potential $V(r)\propto r^\alpha$ yields a mean density $\bar{n}(r)\propto r^{\alpha-2}$, for $\kappa=0$.
Fig.~\ref{fig:densityConf}a. shows results for a quartic potential, $V_4(r)=c_4\, r^4/4$, where $\bar{n}_0(r)\propto r^{2}$. Here a very large 
density step is formed which, together with the radial density increase, further enhances the shell formation from the edge, as for the harmonic confinement, cf. Fig.~\ref{fig:density}a.,d. Next, consider a (regularized) linear confinement, $V_1(r)=c_1[r^2_1+r^2]^{1/2}$, for which 
$\bar{n}(r)\propto r^{-1}$. While again the first density maximum emerges at the boundary, strong modulations of $n(r)$ appear near $r=0$ at low temperatures, Fig.~\ref{fig:densityConf}b, i.e. crystallization proceeds simultaneously from the edge and from the center.

Finally, we further modify the confinement by making the central part of the trap, $0\le r\le r_w$, unaccessible for the particles by using an infinite 
wall at $r=r_w$ together with a shifted harmonic potential ${\tilde V}_2(r)\sim (r-r_w)^2$, cf. inset of Fig.~\ref{fig:densityConf}d. The results for the cases of Coulomb and Yukawa interaction ($\kappa a=1$) are strikingly different, cf. Fig.~\ref{fig:densityConf}c,d. While in the former case again shell formation starts at the edge, in the latter the first shell clearly emerges in the core. 
The reason is that for Coulomb interaction, particles at $r=r_w$ experience almost no radial force, whereas for Yukawa interaction charges located outside do produce an inward force \cite{Christian2006} causing strong particle accumulation at the wall. Not only does this allow to reverse the spatial crystallization 
dynamics, this is also a situation where qualitatively different behavior should be observable for spherically trapped ions (Coulomb interaction) and dusty plasmas (screened interaction).

\emph{In Summary,} we have studied the transition from a weakly coupled to a strongly coupled state in a spherically trapped dusty plasma in a scenario 
which can be realized experimentally. The initial relaxation phase, $\omega_0 t\lesssim 10$, is characterized by formation of a density step $\Delta n$ at the edge and, for low friction, excitation of a breathing oscillation with a $\kappa-$dependent frequency. Around the time $\omega_0 t \sim 10$ shell formation starts at the plasma edge which is a correlation effect arising from the finite density step.
For typical dusty plasma experiments with $\kappa a\approx 0.6$ \cite{bonitz2006} and harmonic confinement, inner shells are formed one by one within almost constant time intervals of $\omega_0t \approx 4$. 
Furthermore, the crystallization dynamics can, to a large extent, be controlled by the shape of the confinement potential. In particular, by 
blocking the central part of the trap crystallization can be initiated in the center.
\begin{acknowledgments}
We thank J.W. Dufty for stimulating comments. This work is supported by the Deutsche Forschungsgemeinschaft via SFB-TR24.
\end{acknowledgments}


\begin{thebibliography}{22}
\bibitem{temperature} We have found that the velocity distribution is close to a Maxwellian which justifies 
the definition of the temperature.
\bibitem{crystal} $\Gamma=125$ is below the value for crystallization which, however, is easily reached if $T_n$ is chosen 
low enough. Then, the radial density modulations further increase and a crystal structure within each shell forms. 
However, this does not change the presented dynamics qualitatively.
\end{thebibliography}

\end{document}